\documentclass[12pt]{article}

\begin{document}

\begin{center}{\Large\bf 
 D(p)-Branes Moving at the}\end{center}
\begin{center}{\Large\bf
Speed of Light}\end{center}

\bigskip

\begin{center}
{\large Bj\o rn Jensen}
\footnote{{\tt e-mail:bjensen@gluon.uio.no}}\footnote{{\tt On leave
from Institute of Physics, University of Oslo, Norway.}}\\
\smallskip
{\it NORDITA,\\ Blegdamsvej 17, DK-2100 Copenhagen \O ,\\ Denmark}
\end{center}

%\begin{center}
%\today
%\end{center}

\bibliographystyle{unsrt}

%\maketitle

%\bigskip

\begin{abstract}
We construct new bosonic boundary states in the light-cone gauge which
describe D(p)-branes moving at the speed of light.
\end{abstract}

\bigskip

\bigskip

\bigskip

\begin{flushleft}
{\bf PACS} number(s): 11.25.-w, 11.25.Hf
\end{flushleft}
%\draft
%\vskip2pc]

\newpage

\section{Introduction}

Static and moving D(p)-branes can either be perceived as spacelike
hyper-surfaces on which open strings can end, or they can equivalently
be described by D(p)-brane boundary states into which closed strings
can disappear. D(p)-branes are characterised by a tension $T_p (\neq
0)$ which can be computed by considering the exchange of closed
strings between two D(p)-branes \cite{Pol}.  Because of this
non-vanishing tension these branes are either static, or they move
with velocities which are less than the speed of light.  A natural
problem to pose, and one which we will pursue in this paper, is
whether it is possible to extend the notion of D(p)-branes above to
also include D(p)-branes which move at the speed of light, or
(equivalently) D(p)-branes which have a vanishing tension $T_p=0$.  In
the following we will call D(p)-branes which move at the speed of
light {\it tensionless} D(p)-branes.

The expression for $T_p$ in terms of the inverse of the fundamental
string tension $\alpha '$, and the string coupling $g$, is given by
\footnote{Note that this expression is true when we are dealing
with static D(p)-branes. When the branes are moving $T_p$ receives
velocity corrections \cite{Arfaei}.}
\begin{equation}
T_p=2\pi g^{-1}(2\pi\sqrt{\alpha '})^{-p-1}\, ,
\end{equation}
where $\mbox{p}\geq 0$ is the number of spacelike directions in the
brane.  When this expression is extrapolated to arbitrarily large
values of $g$, $T_p$ can attain values which are arbitrarily close to
zero.  Hence, in the strongly coupled regime D(p)-branes become light
states (since their masses $M_p$ behaves as $M_p\sim g^{-1}$), and
when $g=\infty$ the D(p)-branes will naively become tensionless.  This
behaviour lies at the very foundation of M(atrix)-Theory
\cite{Susskind}.  However, we will not address the behaviour of
D(p)-branes at infinitely strong coupling in this paper.  We will at
the outset assume that the string coupling constant is relatively
small, so that our considerations are constrained to the domain of
standard string theory.  The question we will address is thus
specifically {\it whether it is possible to construct boundary states
in a perturbative sector of M-Theory which can fit the role as
tensionless D(p)-branes.}

Boundary states which describe moving D(p)-branes were derived in
\cite{Klebanov,Paolo} by acting with a Lorentz boost on the static
boundary states which were derived in \cite{Callan}. This method of
constructing moving D(p)-brane boundary states is necessarily
restricted to the construction of boundary states which describe
D(p)-branes which move with velocities which are less than the speed
of light. In the present paper we demonstrate that the moving bosonic
D(p)-brane boundary states in \cite{Klebanov,Paolo} can be constructed
via a route which is somewhat different from the one followed in those
works.  Our way of deriving these boundary states also allows us to
construct bosonic tensionless D(p)-brane boundary states.
 
Tensionfull D(p)-branes also appear in the super-symmetric type IIA
and type IIB string theories. The bosonic boundary states which we
construct in this paper will also appear in these super-symmetric
theories, and will represent the bosonic pieces of the complete
boundary states which describe D(p)-branes there. The construction of
the complete boundary states which describe tensionless D(p)-branes in
the super-symmetric string theories is straightforward, and will be
presented elsewhere.

We have organised the rest of this paper as follows.  We next rederive
the moving bosonic D(p)-brane boundary states in \cite{Klebanov,Paolo}
(which we will call timelike D(p)-branes), and discuss briefly the
conditions set by the requirement of BRST invariance of bosonic
D(p)-brane boundary states in general. In the third section we
construct boundary states which describe D(p)-branes which move at the
speed of light.  In section four we briefly touch upon the issue about
the nature of the zero-modes of tensionless D(p)-brane boundary
states.  We summarise our results in the last section. The exposition
of the subject matter in this paper is, partially for the purpose of
future reference, aimed at displaying the steps in the various
derivations in the paper in considerable detail.

\section{Timelike D(p)-Branes}

We will first rederive the moving bosonic D(p)-brane boundary states
in \cite{Klebanov,Paolo}.  The boundary state in question is assumed
to move in the $X^{(p+1)}$ target-space direction.  We will
furthermore assume that the classical position of the D(p)-brane in
the spatial directions perpendicular to the brane, is at the origin of
the global coordinate system.  The coordinates perpendicular to the
brane, except for the target-space time-coordinate $X^{(0)}$ and the
$X^{(p+1)}$- coordinate, are
\begin{eqnarray}
{\cal A}=\{ X^{(A)};A\in\{ p+2,...,25\}\}\, ,\nonumber
\end{eqnarray}
and the spacelike coordinates in the brane are
\begin{eqnarray}
{\cal B}=\{ X^{(A)};A\in\{ 1,...,p\}\}\, .\nonumber
\end{eqnarray}
Introduce two new coordinates $Z$ and $Y$ defined by
\begin{equation}
\left(\begin{array}{c} Z\\ Y\end{array}\right)
=\left(\begin{array}{cc} 1 & -v\\ -v &
1\end{array}\right)\left(\begin{array}{c} X^{(p+1)}\\
X^{(0)}\end{array}\right)\, ,
\end{equation}
where $v$ will turn out to be the constant physical linear velocity of
the state ($|v|\leq 1$).  The $Z$-coordinate has been used in previous
works in the context of open strings \cite{Bachas,Lif}. Our choice of
the $Z$ and $Y$ coordinates is guided by the fact that they will give
rise to the correct form of the zero-modes below.  The vectors
associated to $Z$ and $Y$ have the properties
\begin{equation}
\left\{\begin{array}{l} \partial_Z^2=-\partial_Y^2= 1-v^2
\equiv\gamma^2\, ,\\ \partial_Z\cdot\partial_Y=0\, .
\end{array}\right.
\end{equation}
Note that the linear transformation in eq.(2) together with a
normalisation of the $\partial_Z$ and $\partial_Y$ operators will
constitute a Lorentz boost.

We decompose the closed string target-space coordinates into the
canonical form
%\newpage
\begin{eqnarray}
&&X^{(A)}(\sigma ,\tau )= x^{(A)}+p^{(A)}\tau +\frac{i}{2}\sum_{n\neq
0}\frac{1}{n}(\alpha^{(A)}_n e^{-2in(\tau -\sigma
)}+\tilde{\alpha}^{(A)}_ne^{-2in(\tau +\sigma )})
\end{eqnarray}
for any $A\in \{ 0,...,25 \}$.  The $\alpha_m$'s are related to the
conventionally normalised harmonic oscillator operators by
$\alpha^{(A)}_m=\sqrt{m}a^{(A)}_m$ and
$\alpha^{(A)}_{-m}=\sqrt{m}a^{(A)\dagger}_m$ ($m>0$).  The, at the
outset, uncoupled $\alpha^{(A)}_m$ and $\tilde{\alpha}^{(A)}_m$
operators are normalised to the flat Minkowski metric $\eta^{AB}$ with
signature $(-++....)$, i.e.
\begin{eqnarray}
&&[\alpha^{(A)}_m,\alpha^{(B)}_n]=[\tilde{\alpha}^{(A)}_m,
\tilde{\alpha}^{(B)}_n]=
m\delta_{m+n}\eta^{AB}\, ,\\ &&[\alpha^{(A)}_m,\alpha^{(B)}_n]=0\, .
\end{eqnarray}

We will consider the following mixed set of Dirichlet and Neumann
boundary conditions on the closed string target-space coordinates
%\newpage
\begin{eqnarray}
\mbox{I}&:&(\partial_\sigma X^{(A)})|_{\tau =0}|B;p,v\rangle =0\, ,\,
X^{(A)}|_{\tau =0}|B;p,v\rangle =0\, ;\, A\in{\cal A}\, ,\\
\mbox{II}&:&(\partial_\tau X^{(A)})|_{\tau =0}|B;p,v\rangle =0 \, ;\,
A\in{\cal B}\, ,\\ \mbox{III}&:&(\partial_\sigma Z)|_{\tau
=0}|B;p,v\rangle = 0\, ,\, Z|_{\tau =0}|B;p,v\rangle =0\, ,\\
\mbox{IV}&:&(\partial_\tau Y)|_{\tau =0}|B;p,v\rangle =0\, .
\end{eqnarray}
$|B;p,v\rangle$ denotes the state-vector of a D(p)-brane boundary
state with intrinsic spatial dimensionality p, and velocity $v$.
$\sigma$ and $\tau$ are the spacelike and timelike coordinates in the
closed string world-sheet, respectively.  We will let lower-case Greek
letters denote the world-sheet coordinates in the following.  We have
also assumed, without any loss of generality, that the closed strings
are absorbed (or emitted) by the boundary state at world-sheet time
$\tau =0$.

The boundary conditions I-IV above translate into the following set of
operator relations when we substitute the expansion in eq.(4) into
eq.(7-10), and equate coefficients of $e^{im\sigma}$
\begin{eqnarray}
\mbox{I}&:& \alpha^{(A)}_{-m}=\tilde{\alpha}^{(A)}_m\, ,\, x^{(A)}=0
\, ;\, A\in{\cal A}\, ,\\ \mbox{II}&:&
\alpha^{(A)}_{-m}=-\tilde{\alpha}^{(A)}_m\,\, ,\,\, p^{(A)}=0\, ;\,
A\in{\cal B}\, ,\\ \mbox{III}&:& \alpha^{(p+1)}_m+
v\alpha^{(0)}_m=\tilde{\alpha}^{(p+1)}_{-m}+v\tilde{\alpha}^{(0)}_{-m}
\, , x^{(p+1)}=vx^{(0)}\, ,\\ \mbox{IV}&:&
v\alpha^{(p+1)}_m+\alpha^{(0)}_m=
-(v\tilde{\alpha}^{(p+1)}_{-m}+\tilde{\alpha}^{(0)}_{-m})\,\, ,\,\,
vp^{(0)}=p^{(p+1)}\, .
\end{eqnarray}
The structure of the resulting operator relation in I and the
resulting operator relation in II is that a right moving creation
operator equals a left moving annihilation operator (say). Clearly,
the operator relations in III and IV also have this structure, but in
order to make this feature more transparent we define the following
new set of operators using an obvious notation
\begin{eqnarray}
A_r\equiv \alpha^{(p+1)}_r+ v\alpha^{(0)}_r\,\, &,&\,\,
\tilde{A}_{-r}\equiv\tilde{\alpha}^{(p+1)}_{-r}+
v\tilde{\alpha}^{(0)}_{-r}
\,\, ; r\neq 0\, ,\\ B_s\equiv v\alpha^{(p+1)}_s+\alpha^{(0)}_s\,\,
&,&\,\, \tilde{B}_{-s}\equiv
v\tilde{\alpha}^{(p+1)}_{-s}+\tilde{\alpha}^{(0)}_{-s} \,\, ; s\neq
0\, .
\end{eqnarray}
In terms of these operators the operator relations in III and IV
become
\begin{equation}
A_r=\tilde{A}_{-r}\,\, ,\,\, B_r=-\tilde{B}_{-r}\, .
\end{equation}
Clearly, we have that $A_{-r}=A^\dagger _r$, $B_{-s}=B^\dagger _s$
($r,s>0$).  The new operators $A$, $B$, $\tilde{A}$ and $\tilde{B}$
can easily be seen to satisfy the following uncoupled, and closed
algebras
\begin{equation}\left\{\begin{array}{l}
[A_r,B_s] =0 \, ,\\

[A_r,A_{s} ] = -[B_r,B_{s} ] =r(1-v^2)\delta_{r+s} \, ,\\
\end{array}\right.\end{equation}
and
\begin{equation}\left\{\begin{array}{l}
[\tilde{A}_r,\tilde{B}_s]=0 \, ,\\

[\tilde{A}_r,\tilde{A}_{s} ]= -[\tilde{B}_r,\tilde{B}_{s}
]=r(1-v^2)\delta_{r+s}\, .
\end{array}\right.\end{equation}
It is evident that these algebras have the same structure as the
algebra in eq.(5-6).  With $v=0$ we find that $A_r= \alpha^{(p+1)}_r$
and $B_s=\alpha^{(0)}_s$. This suggests that we in general (provided
that $|v|<1$, of course) can interpret the $A_r$ operators as
corresponding to spacelike excitations, while the $B_s$ operators can
be looked upon as timelike excitations.  The same reasoning also holds
in the tilded sector.  Clearly, the signs of the commutators in
eq.(18) and eq.(19) are consistent with our interpretation of these
operators.  The algebra in eq.(18) has the same structure as the
algebra in eq.(19). Hence, the right moving sector of operators is
just a copy of the left moving sector of operators on this algebraic
level.

%\smallskip

The way we have organised the operators in eq.(11), eq.(12) and
eq.(17) suggests to use the operators $\alpha_{m}^{(A)}$ with $A\neq
(0, p+1)$, as well as the $A_r$ and $B_s$ operators as a basis for a
canonical coherent state construction of the moving D(p)-brane
boundary states.  In a canonical coherent state construction which
involves an infinite number of bosonic degrees of freedom, one assumes
the existence of an infinite set of annihilation and creation
operators $d_k$, $d_{-k}$ ($k,l:$ positive integers) which are
normalised to $(\pm )$ unity (i.e., $[d_k,d_{-l}]=\pm\delta_{k-l}$),
and the existence of a vacuum state $|0\rangle\in{\cal F}$ on which
the creation operators can act so as to produce the usual Fock space
representation of the canonical commutation relations.  Let $\{ z_k\}$
denote an arbitrary sequence of complex c-numbers such that $\sum_{k}
|z_k|^2 <\infty$.  $\{ z_k\}$ defines an element in an infinite
dimensional Hilbert space ${\cal H}$.  To each such element in ${\cal
H}$ we define a unit vector $|\{ z_k\}\rangle\in {\cal F}$.  One can
then construct an over-complete set of basis vectors for ${\cal F}$
with each of the vectors given by
\begin{equation}
|\{ z_k\}\rangle = e^{(-\frac{1}{2}\sum_{k=1}^{\infty}|z_k|^2)} e^{(
\sum_{k=1}^{\infty}\pm z_kd_{-k})} |0\rangle\, ,
\end{equation}
where we must use $+$ in the last sum when $d_k$ is normalised to
unity, while we must use the minus sign when this operator is
normalised to minus one.  A key property of the canonical coherent
states is that
\begin{equation}
d_k|\{ z_k\}\rangle =z_k|\{z_k\}\rangle\, .
\end{equation}

In order to go through with a coherent state construction of
D(p)-brane boundary states, we may construct the boundary states using
right-moving operators, e.g..  Seen from the perspective of the
right-moving sector left-moving operators appear as c-numbers, and
will thus naturally play the role as the $\{ z_k\}$-sequences
above. We can of course swap the roles played by the right and
left-moving sectors in the construction of the boundary states. It is
thus in some sense natural and ``democratic'' to normalise the
$\alpha^{(A)}_m$ $(A\neq 0,p+1)$, $A_r$ and $B_s$ operators, as well
as their eigenvalues $\tilde{\alpha}^{(A)}_{-m}$, $\tilde{A}_{-r}$ and
$\tilde{B}_{-s}$, to unity, i.e. we should rescale these operators
according to ($m,r,s>0$)
\begin{equation}
\alpha^{(A)}_m\rightarrow\frac{1}{\sqrt{m}}\alpha^{(A)}_m\,\, ,\,\,
A_r\rightarrow \frac{1}{\sqrt{r}\gamma}A_r \,\, ,\,\, B_s\rightarrow
\frac{1}{\sqrt{s}\gamma}B_s \,\, ,\,\,
\end{equation}
\begin{equation}
\tilde{\alpha}^{(A)}_{-m}\rightarrow\frac{1}{\sqrt{m}}
\tilde{\alpha}^{(A)}_{-m}\,\,
,\,\,
\tilde{A}_{-r}\rightarrow\frac{1}{\sqrt{r}\gamma}\tilde{A}_{-r}\,\,
,\,\,
\tilde{B}_{-s}\rightarrow\frac{1}{\sqrt{s}\gamma}\tilde{B}_{-s}\, .
\end{equation}
This corresponds to introducing conventionally normalised harmonic
oscillators $a^{(A)}_m$, as well as normalising the vectors
$\partial_Z$ and $\partial_Y$ to unity. With this rescaling we find
that the oscillator part of a moving bosonic D(p)-brane boundary state
is given by
%\newpage
\begin{eqnarray}
&&|B;p,v\rangle \sim
exp(+\sum_{m=1}^{\infty}\frac{1}{m}(\sum_{A=p+2}^{25}\alpha^{(A)}_{-m}
\tilde{\alpha}^{(A)}_{-m}-
\sum_{D=1}^{p}\alpha^{(D)}_{-m}\tilde{\alpha}^{(D)}_{-m}+\nonumber\\
&&+\frac{1}{\gamma^2}(A_{-m}\tilde{A}_{-m}+B_{-m}\tilde{B}_{-m})))
|0\rangle\otimes |B;p\rangle_{\mathrm{ghost}} =\nonumber\\
&&=exp(+\sum_{m=1}^{\infty}(\sum_{A=p+2}^{25}a^{(A)\dagger}_{m}
\tilde{a}^{(A)\dagger}_{m}-
\sum_{D=1}^{p}a^{(D)\dagger}_{m}\tilde{a}^{(D)\dagger}_{m}))
\times\nonumber\\
&&\times exp (+\sum_{m=1}^{\infty} (a^{(0)\dagger}_{m}\, ,\,
a^{(p+1)\dagger}_{m}){\mbox{\bf M}} (v) \left(\begin{array}{c}
\tilde{a}^{(0)\dagger}_{m}\\
\tilde{a}^{(p+1)\dagger}_{m}\end{array}\right)) |0\rangle\otimes
|B;p\rangle_{\mathrm{ghost}}\, ,
\end{eqnarray}
where we in the last two lines have switched to the conventionally
normalised harmonic oscillator operators in order to facilitate the
comparison with previous results.  In eq.(24) $|0\rangle$ denotes the
bosonic closed string tachyon vacuum, $|B;p\rangle_{\mathrm{
ghost}}$ denotes the ghost part of the full boundary state, and the
matrix $\mbox{\bf M}(v)$ is given by
\begin{equation}
{\mbox{\bf M}}(v)= \left(\begin{array}{cc} \frac{1+v^2}{1-v^2} &
\frac{+2v}{1-v^2}\\ \frac{+2v}{1-v^2} & \frac{1+v^2}{1-v^2}
\end{array}\right)=\frac{1}{\gamma^2}\left(\begin{array}{cc}
1+v^2 & +2v\\ +2v &
1+v^2\end{array}\right)\equiv\frac{1}{\gamma^2}{\mbox{\bf L}}(v)\, .
\end{equation}
In the expression for the boundary states we have omitted an overall,
but for our purposes, unimportant normalisation constant, as well as
neglected the contribution from the zero-modes.  We return to the form
of the zero-modes near the end of this paper.  When we deal with a
D(p)-brane boundary state which has no spatial extension, i.e. a
D-particle, the sum over $D$ is absent in eq.(24).  The quantum states
which are described by eq.(24) are exactly equal to the bosonic
D(p)-brane boundary states which were constructed in
\cite{Paolo}. Note that the letter $v$ in \cite{Paolo} denotes the
{\it rapidity} of the boundary state.

When we follow the prescription for the construction of a bosonic
coherent (boundary) state we are, strictly speaking, only forced to
normalise the untilded (say) sector, i.e. we must either enforce
eq.(22) {\it or} eq.(23) but we must not necessarily enforce both sets
of rescalings.  If we only impose eq.(22) or eq.(23), the boundary
state will still have the same form as above, but with ${\mbox{\bf
M}}(v)$ given by ${\mbox{\bf M}}(v)=\gamma^{-1}{\mbox{\bf L}}(v)$. We
have ({\it a priori}) no other arguments at this level other than
``naturalness'', for why we should choose ${\mbox{\bf M}}(v)$ to have
the form in eq.(25) rather than the alternative expression for this
matrix.

Let us briefly look into the question of BRST invariance of D(p)-brane
boundary states in some generality.  The BRST charge in the untilded
sector is defined by
\begin{equation}
{\cal Q}
=\sum_{n=-\infty}^{\infty}:L_{-n}c_n:-\frac{1}{2}\sum_{m,n=-\infty}
^{\infty}(m-n):c_mc_nb_{m+n}:-c_0\, ,
\end{equation}
with a corresponding expression in the tilded sector. :: denotes
normal ordering of the operators.  The $c_n$ and $b_n$ operators
represent the ghost and anti-ghost operators respectively, and the
$L_n$'s are the Virasoro operators defined by
\begin{equation}
L_m=\frac{1}{2}\sum_{n=-\infty}^{\infty}\alpha^{(A)}_{m-n}
\alpha_{(A)n}\, .
\end{equation}
The BRST invariance condition is simply
\begin{equation}
({\cal Q}+\tilde{\cal Q})|B;p,v\rangle =0\, .
\end{equation}
The nil-potency of the BRST charge holds, of course, by construction,
since we are working with critical bosonic string theory.
 
Let us consider the BRST invariance of the {\it static} boundary
states $|B;p,0\rangle$. From the first part in the BRST charge above
(which contains the Virasoro operators), and the corresponding part in
the tilded sector, we get the condition
\begin{equation}
\frac{1}{2}\sum_{m,n=-\infty}^{\infty}:\tilde{\alpha}_{-(m+n)}^
{(A)}\tilde{\alpha}_{n(A)}:
(c_m+\tilde{c}_{-m})|B;p,0\rangle=0\,\, ;\,\, A\in\{0,...,25\}
\end{equation}
after having imposed the boundary conditions in eq.(11) and
eq.(12). Hence, we find that we must set
\begin{equation}
c_m=-\tilde{c}_{-m}
\end{equation}
in order to make the expression in eq.(29) vanish. Similarly, the
second term in the BRST charge implies that (after having imposed the
conditions in eq.(11) and eq.(12))
\begin{equation}
-\frac{1}{2}\sum_{m,n=-\infty}^{\infty}(m-n):(\tilde{c}_m
\tilde{c}_n\tilde{b}_{m+n}-
\tilde{c}_m\tilde{c}_nb_{-(m+n)}):|B;p,0\rangle =0\, ,
\end{equation}
such that we likewise find that we must impose
\begin{equation}
\tilde{b}_m=b_{-m}\, .
\end{equation}
The last term in the BRST charge does not give rise to any additional
conditions.  The conditions in eq.(30) and in eq.(32) were also
derived in \cite{Callan}.  Clearly, these conclusions are invariant
under Lorentz transformations.  Since the moving boundary states
$|B;p,0<v<1\rangle$ can be obtained from the static states via a boost
transformation in the $\mbox{p}+1$ target-space direction, it follows
that these moving states are physically acceptable BRST invariant
states.
\footnote{Clearly, since the ghost fields do not carry Poincare
indices the $|B;p\rangle_{\mathrm{ghost}}$ states have the same
form for all values of $v$. Hence, the ghost part of a boundary state
does not carry a specific velocity label. The author thanks
Prof. Paolo Di Vecchia for this simple argument.}  It also follows
that we must choose $\mbox{\bf M}(v)$ as in eq.(25), and not the
alternative expression for this matrix, since the form of this matrix
in eq.(25) also follows from boost transforming the static boundary
states.

Let us now probe the possibility to construct D(p)-brane boundary
states $|B;p,1\rangle$ which move at the speed of light from the
tensionfull states. The timelike states $|B;p,|v|<1\rangle$ are not
defined in the limit when $|v|\rightarrow 1$ due to the normalisation
factor $\gamma$ which we introduced in eq.(22) and eq.(23). {\em
Hence, we cannot derive tensionless states $|B;p,1\rangle$ by just
taking the naive $|v|\rightarrow 1$ limit of the timelike D(p)-brane
states.}  On a more fundamental level it is clear that the tensionfull
and tensionless D(p)-brane sectors are well separated, since when we
set $|v|=1$ we have that $A_r=B_r$ and $\tilde{A}_r=\tilde{B}_r$,
which also implies that the algebras in eq.(18) and eq.(19)
degenerate. This algebraic degeneration reflects the fact that the
conditions on the oscillators which stem from eq.(9) and eq.(10)
almost coincide on the light-cone.  It is also clear that eq.(17)
implies that we cannot impose both eq.(9) and eq.(10) when $|v|=1$,
i.e. not both of these conditions can be continued to the light-cone.
We furthermore note another obstruction to a ``naive'' construction of
light-like boundary states: $A_{r}$, $A_{-r}$, and $\tilde{A}_{r}$,
$\tilde{A}_{-r}$ {\it commute} when $|v|=1$.  It follows that these
operators cannot be used in a canonical coherent state construction of
tensionless D(p)-brane boundary states, since such a scheme
presupposes that the operators are normalised to unity.

When $|v|=1$, and after having imposed the conditions in eq.(11),
eq.(12), eq.(30) and eq.(32), we find that the BRST condition in
eq.(28) is reduced to
\begin{equation}
({\cal Q}+\tilde{\cal Q})|B;p,1\rangle
=\frac{1}{2}\sum_{m=-\infty}^{\infty}:(\tilde{l}_m-
{l}_{-m})\tilde{c}_m:|B;p,1\rangle =0\, .
\end{equation}
In this equation we have introduced the new operators $l_m$ and
$\tilde{l}_m$ which have the same form as the Virasoro operators
above, but where the target-space index only runs over $0$ and
$\mbox{p}+1$. Hence, we have to impose the following extra constraint
on the $|B;p,1\rangle$ states
\begin{eqnarray}
%\left\{\begin{array}{c}
l_m|B;p,1\rangle =\tilde{l}_{-m}|B;p,1\rangle\, .
\end{eqnarray}
In the next section we will turn to the light-cone gauge in which all
our problems sofar with the construction of tensionless D(p)-brane
boundary states will vanish.

\section{Tensionless D(p)-Branes}

When one deals with D(p)-branes which move at the speed of light it is
natural to consider the construction of the corresponding boundary
states in an adapted frame, i.e.  in the light-cone gauge.  In the
following we will use the letters $\tau$ and $\sigma$ to denote the
world-sheet coordinates also in the light-cone gauge.  The light-cone
coordinates $X^{(\pm )}$ are defined by
\begin{equation}
X^{(\pm )}=\frac{1}{\sqrt{2}}(X^{(0)}\pm X^{(p+1)})\, .
\end{equation}
We choose $X^{(+)}$ as the ``time-coordinate'' such that we
consistently can set all the $\alpha^{(+)}_m$ and
$\tilde{\alpha}^{(+)}_m$ oscillators to zero. $X^{(+)}$ will thus
correspond to the $Y$ coordinate in the previous section.  We then
have that $X^{(+)}=x^{(+)}+p^{(+)}\tau$, where $p^{(+)}$ is the
light-cone energy.  The $X^{(-)}$-coordinate is, from the string
equations of motion, constrained to satisfy the relation
\begin{equation}
p^{(+)}\partial_\sigma X^{(-)}=\partial_\tau X^{(I)}\partial_\sigma
X^{(I)} \, ;\, I\in {\cal A}\cup {\cal B}\, .
\end{equation}
Hence, eq.(7) and eq.(8) imply that $X^{(-)}$ should satisfy the
Dirichlet condition
\begin{equation}
\partial_\sigma X^{(-)}|_{\tau =0}|B;p,1\rangle =0\, .
\end{equation}
This condition coincides with the condition in eq.(9). It did also
appear in \cite{Gutperle} in their discussion of static D(p)-brane
boundary states in type II super-string theory in the light-cone
gauge.  In order to constrain the zero-modes to lie on the light-cone,
we also impose the second constraint in eq.(9), i.e.
\begin{equation}
X^{(-)}|_{\tau =0}|B;p,1\rangle =0\, ,
\end{equation}
i.e. the $X^{(-)}$-coordinate plays the same role as the
$Z$-coordinate in the previous section.  When we insert the expansion
in eq.(4) into eq.(37) and eq.(38) and equate coefficients of
$e^{im\sigma}$, we find that ($m\neq 0$)
\begin{eqnarray}
&&\alpha^{(-)}_m=\tilde{\alpha}^{(-)}_{-m}\, ,\\
&&x^{(-)}=x^{(0)}-x^{(p+1)}=0\, .
\end{eqnarray}
In the light-cone gauge there are no ghosts manifestly present.  In
order to construct tensionless D(p)-brane boundary states the relation
in eq.(39) is thus the only additional constraint on the oscillators
we have to take into account in addition to the constraints in eq.(11)
and eq.(12).

Note that the boundary states $|B;p,1\rangle$ are positioned at a
fixed ``time'' $X^{(+)}$ since
\begin{equation}
X^{(+)}|_{\tau =0}|B;p,1\rangle = x^{(+)}= \mbox{constant}\,\, ,\,\,
\partial_\sigma X^{(+)}|_{\tau =0}|B;p,1\rangle =0\, .
\end{equation}
The resulting kinematics really describe $(\mbox{p}+1)$-instanton
states rather than D(p)-brane states, since $X^{(+)}$ satisfies a
Dirichlet boundary condition \cite{Gutperle}.  The D(p)-brane states
are related to the instanton states by a double Wick rotation
\cite{Gutperle}.  When we use the words ``D(p)-brane'' in the
following it is really this rotated version we are referring
to. Clearly, mapping $X^{(+)}$ on a lightlike circle of radius $R$,
the dual lightlike circle has radius $\alpha '/R$.  Let the dual
circle be coordinatised by the coordinate $X^{(+)}_{\mathrm{
Dual}}$.  The general relation $\partial_\mu
X^{(A)}=\epsilon_{\mu\nu}\partial^\nu X^{(A)}_{\mathrm{Dual}}$
then implies that
\begin{equation}
\partial_\sigma X^{(+)}|_{\tau =0}|B;p,1\rangle =0 \Rightarrow
\partial_\tau X^{(+)}_{\mathrm{Dual}}|_{\tau =0}|B;p,1\rangle
=0\, .
\end{equation}
Hence, in the dual space the boundary states satisfy a set of boundary
conditions which are {\em exactly} analogous to the ones in eq.(7-10).

In the light-cone gauge we only deal with the transversal degrees of
freedom. It follows that the condition in eq.(39) can be expressed
entirely in terms of the $\alpha^{(I)}_m$ operators, which at the
outset satisfy eq.(11) and eq.(12).  Hence, the relation in eq.(39)
may potentially induce further relations among, or conditions on, the
transversal oscillators. However, since we from eq.(36) have that
\begin{equation}
\alpha^{(-)}_m=\frac{1}{2p^+}(\sum_{I}\sum_{r=-\infty}^{\infty}:
\alpha^{(I)}_{m-r}\alpha^{(I)}_r:-\delta_m)\, ,
\end{equation}
\begin{equation}
\tilde{\alpha}^{(-)}_m=\frac{1}{2p^+}(\sum_{I}\sum_{r=-\infty}^{\infty}:
\tilde{\alpha}^{(I)}_{m-r}\tilde{\alpha}^{(I)}_r:-\delta_m)\, ,
\end{equation} 
it is easy to show that eq.(39) reduces to an identity relation when
we impose eq.(11) and eq.(12) on the right or the left hand side of
eq.(39).

It is now straightforward to construct an explicit expression for the
oscillator part of the tensionless D(p)-brane boundary states in the
light-cone gauge, since they are only determined by the boundary
conditions in eq.(11) and eq.(12).  It follows that these states then
can be written as
\begin{equation}
|B;p,1\rangle\sim
exp(+\sum_{m=1}^{\infty}(\sum_{A=p+2}^{25}a^{(A)\dagger}_{m}\tilde{a}^
{(A)\dagger}_{m}- \sum_{D=1}^{p}a^{(D)\dagger}_{m}\tilde{a}^
{(D)\dagger}_{m}))|0\rangle\, ,
\end{equation}
where we again have switched to conventionally normalised operators.

\section{Zero-Modes}

We have so far in our treatment not discussed how we should treat the
zero-modes of the D(p)-brane boundary states.  The zero-modes describe
the classical position of a D(p)-brane. Naively we should therefore
expect that it is sufficient to include $\delta$-functions in the
expressions for the boundary states in order to take these modes
properly into account. Hence, the $|B;p,|v|<1\rangle$ states should
naively be multiplied with
\begin{equation}
N_p\delta (x^{(p+1)}-vx^{(0)})\prod_{J\in{\cal A}}\delta (x^{(J)})\, ,
\end{equation}
where $N_p$ is an overall normalisation constant. The tensionless
states should likewise be multiplied with a similar kind of
expression, but with $v$ set to unity, i.e.
\begin{equation}
M_p\delta (x^{(-)})\prod_{J\in{\cal A}}\delta (x^{(J)})\, ,
\end{equation} 
where $M_p$ is another overall normalisation constant.  However, in
\cite{Paolo} it was emphasised that eq.(46) is not the completely
correct expression, since a Lorentz boost on $\delta (x^{(p+1)})$
results in $\sqrt{1-v^2}\delta (x^{(p+1)}-vx^{(0)})$, i.e. a
Born-Infeld type factor should also be included in the complete
specification of the boundary states.  That the extra multiplicative
term indeed is the effective Born-Infeld action for the D(p)-brane in
question was shown in \cite{Klebanov}.  Hence, in order to derive the
complete expression for the zero-mode part of the tensionless boundary
states, we should probably first derive the effective action for
tensionless D(p)-branes.  This will not be attempted in this paper.

\section{Summary and Conclusion}

Our way of constructing tensionfull D(p)-brane boundary states is
divided into two steps:
\begin{flushleft}{\bf (1)} the choice of coordinates
in eq.(2) together with the set I-IV of boundary conditions in
eq.(7-10), and\end{flushleft}
\begin{flushleft}{\bf (2)} the set
of rescalings in eq.(22) and eq.(23) which are necessary to impose in
order to represent D(p)-brane boundary states as canonical coherent
states.\end{flushleft} We noted the existence of an ambiguity in the
necessary number of rescalings, which made the particular choice of
the form of the matrix $\mbox{\bf M}$ in eq.(24) somewhat unclear.
The purpose of the reanalysis of the construction of moving
tensionfull D(p)-brane boundary states was to see in detail what
``goes wrong'' when one attempts to extend the notion of D(p)-branes
to also cover the case when these states move at the speed of light.
Clearly, step {\bf (2)} above cannot be part of such a scheme due to
diverging normalisation factors.  However, even though one overlooks
this problem (since the normalisation is linked to the wish of
representing the states as {\em canonical} coherent states) it is
nevertheless inconsistent to try to impose the set I-IV in eq.(7-10)
directly when $|v|=1$.  It was specifically noted that it is
inconsistent, due to eq.(17), to impose both the condition III and the
condition IV in eq.(9-10) at the same time when $|v|=1$.

In section 3 we attacked the problem of constructing D(p)-brane
boundary states from a radically different perspective than the one
used in the previous section.  We showed that imposing only the
conditions in I, II and III (in eq.(7-9)) leads to a set of boundary
conditions which are consistent with the string equations of motion in
{\em the light-cone gauge}, and which after a duality transformation
describe D(p)-brane boundary states which move at the speed of light.
{\em Hence, we discovered that the tensionless states in the dual
space are completely determined by, and can be constructed entirely
from, a set of boundary conditions which are exactly analogous to the
ones in I-IV in eq.(7-10).  In this way we managed to show that it is
possible to derive boundary states which describe D(p)-branes moving
at {\em any} physical velocity in a ``unified'' manner.}

The derivation of, as well as the actual expression in eq.(45) for,
the tensionless D(p)-brane states are remarkably simple. We note in
particular that the explicit expressions for the $|B;p,1\rangle$
vectors are (implicitly) independent of the $X^{(0)}$ and
$X^{(p+1)}$-oscillators. It follows that the formal expressions for
the tensionless states are unaltered under any duality transformation
involving the $X^{(0)}$ and $X^{(p+1)}$-coordinates. This points to a
crucial difference between tensionless and tensionfull D(p)-branes
when we compare the behaviour of the $|B;p,|v|<1\rangle$ states under
T-duality with the corresponding properties of tensionless
D(p)-branes.  In the dual expression for the tensionfull states the
momentum carried by a moving D(p)-brane is transmuted into an electric
field (in a static D(p+1)-brane) in the dual (super-) Yang-Mills
description when we dualise on the $X^{(p+1)}$-coordinate.  When we
let $|v|\rightarrow 1$ this electric field becomes super-critical, and
it is expected to decay via the usual quantum mechanical mechanisms
for decay of strong fields \cite{Bachas}.  This {\em prohibits on
physical grounds} the possibility to boost a $|B;p,|v|<1\rangle$ state
into a (near) $|B;p,1\rangle$ state. Correspondingly, we pointed out
that it is {\em inconsistent on an algebraic level} to try to smoothly
relate a $|B;p,|v|<1\rangle$ state to a tensionless D(p)-brane
state. Hence, from the tensionfull D(p)-brane side both physical as
well as more formal algebraic arguments exist which together
effectively prohibits a direct and smooth relation between tensionfull
and tensionless D(p)-brane boundary states. The corresponding argument
from the tensionless D(p)-brane side is the observation above that the
states in eq.(45) are insensitive to any duality transformation in the
$X^{(0)}$ and $X^{(p+1)}$ directions. Hence, a $|B;p,1\rangle$ state
cannot in particular be related to a static $D(p+1)$-brane via
duality.

Let us conclude this paper. We have shown that tensionless D(p)-brane
states exist in string theory. This tensionless sector is analogous to
the massless sector in point-particle theories, and it is effectively
separated from the tensionfull D(p)-brane sector. We plan to return to
a number of questions which are raised by this work together with the
construction of tensionless D(p)-brane boundary states in super-string
theory elsewhere. Of particular further interest is to 
understand the role played by tensionless D(p)-branes in string
perturbation theory, and to construct and
investigate the super-Yang-Mills theory which may be associated to
tensionless D(p)-branes in super-string theory.

\section{Acknowledgements}

The author thanks Dr. Marco Billo, Dr. Daniel Cangemi, Dr. Olav
 Tirkonnen and Prof. Paolo Di Vecchia for valuable discussions, and
Dr. M.M. Sheikh Jabbari for making him aware of ref.\cite{Arfaei}.  He
 also thanks Prof. I. Brevik for reading and commenting upon a near
 final version of the paper, and Prof. C. Bachas,
Prof. M.B. Green and Prof. L. Thorlacius for 
interesting and encouraging comments on the first 
web-version of this paper.
The author acknowledges the generous
 hospitality of NORDITA and The Niels Bohr Institute during the time
 this work was carried out.

%\newpage

\end{document}